\documentclass
[reprint,aps,prapp,letter,twocolumn,english,superscriptaddress,showpacs,10pt]{revtex4-2}%
\usepackage{eurosym}
\usepackage{amsmath,bm}
\usepackage{mathtools}
\usepackage{commath}
\usepackage{amssymb}
\usepackage{graphicx}
\usepackage{babel}
\usepackage{cases}
\usepackage[colorlinks=true, citecolor=blue, anchorcolor=green]{hyperref}
\usepackage{natbib}
\usepackage{floatrow}
\usepackage{xcolor}
\usepackage{braket}
\usepackage{lipsum}%
\usepackage{amsmath}%
\usepackage{amsfonts}
\providecommand{\U}[1]{\protect\rule{.1in}{.1in}}
\hypersetup{linkcolor=red}
\makeatletter

\newcommand{\Rmnum}[1]{\expandafter\@slowromancap\romannumeral #1@}

\makeatother
\begin{document}

\title{A Perspective on Ferrons}

\author{G.E.W. Bauer}
\affiliation{WPI Advanced Institute for Materials Research and CSRN, Tohoku University, 2-1-1, Katahira, Sendai 980-8577, Japan}
\affiliation{Institute for Materials Research, Tohoku University, 2-1-1 Katahira, Sendai 980-8577, Japan}
\affiliation{Kavli Institute for Theoretical Sciences, University of the Chinese Academy of Sciences, Beijing 10090, China}

\author{P. Tang}
\affiliation{WPI Advanced Institute for Materials Research and CSRN, Tohoku University, 2-1-1, Katahira, Sendai 980-8577, Japan}

\author{R. Iguchi}
\affiliation{National Institute for Materials Science, Tsukuba 305-0047, Japan}

\author{J. Xiao}
\affiliation{Department of Physics and State Key Laboratory of Surface Physics, Fudan University, Shanghai 200433, China}

\author{K. Shen}
\affiliation{Center for Advanced Quantum Studies and Department of Physics, Beijing Normal University, Beijing 100875, China}

\author{Z. Zhong}
\affiliation{Ningbo Institute of Materials Technology and Engineering, Chinese Academy of Sciences, Ningbo 315201, China} \affiliation{China Center of Materials Science and Optoelectronics Engineering, University of the Chinese Academy of Sciences, Beijing 100049, China}

\author{T. Yu}
\affiliation{School of Physics, Huazhong University of Science and
Technology, Wuhan 430074, China}

\author{S.M. Rezende}
\affiliation{Departamento de F\'isica, Universidade Federal de Pernambuco,
50670-901, Recife, Pernambuco, Brazil}

\author{J.P. Heremans}
\affiliation{Department of Materials Science and Engineering, The Ohio State
University, Columbus, Ohio 43210, USA} \affiliation{Department of Mechanical
and Aerospace Engineering, The Ohio State University, Columbus, Ohio 43210,
USA} 
\affiliation{Department of Physics, The Ohio State University,
Columbus, Ohio 43210, USA}

\author{K. Uchida}
\affiliation{National Institute for Materials Science, Tsukuba 305-0047,
Japan} 
\affiliation{Institute for Materials Research, Tohoku University, 2-1-1 Katahira, 980-8577 Sendai, Japan}

\date{\today }

\begin{abstract}
The duality between electric and magnetic dipoles in electromagnetism only partly applies to condensed matter. In particular, the elementary excitations of the magnetic and ferroelectric orders, namely magnons and ferrons, respectively, have received asymmetric attention from the condensed matter community in the past. In this perspective, we introduce and summarize the current state of the budding field of "ferronics" and speculate about its potential applications in thermal, information, and communication technology.

\end{abstract}

\maketitle

\section{Ferroelectricity and magnetism}

Ferroelectrics are materials that exhibit permanent electric polarization resulting from the ordering of electric dipoles below a high critical temperature. Comparing and contrasting the behavior of ferroelectrics with that of magnets, in which magnetic dipoles order, can enhance our understanding of both material classes \cite{Spaldin, Duality}. Due to their high dielectric constant, ferroelectrics are commonly used in (power) capacitors. Their polarization can be manipulated by external electric fields and temperature, leading to applications such as electrocaloric cooling via thermodynamic cycles similar to those used in magnetocalorics \cite{Crossley}. Pyro- and piezoelectricity can be utilized for sensor and energy harvesting applications \cite{Kang}. The finite coercivity of the switchable order makes ferroelectrics useful materials for non-volatile memories that compete with magnetism-based devices \cite{Memories}.

The polarization of a polar material is affected by the details of its surface, but the distinction between a ferroelectric and an unpolarized dielectric is a well-defined bulk property that can be computed from first principles \cite{Vanderbilt}. While the analogies between ferroelectricity and magnetism are well-understood for equilibrium thermal properties in terms of the entropy of the dipolar order, much less attention has been given to the excited states of ferroelectrics, that we dubbed "ferrons" \cite{Bauer2}, compared to the corresponding states in magnets, known as "magnons" \cite{MagnonicsTB, Magnonics}. In this Perspective, we summarize the state-of-the-art of ferronics, with a particular focus on its potential applications.

\section{Ferroelectric materials}
The exchange interaction drives magnetism by favoring parallel spins over antiparallel ones. The magneto-dipolar interaction is relatively weak and causes secondary effects, such as the formation of domains. In ferroelectrics, there is no exchange interaction but the dipolar interaction is stronger by a factor of  \(1/(\alpha^2 \epsilon_r) \approx 10^3\), where $\alpha=1/137$ is the fine-structure constant and \(\epsilon_r\) is a high-frequency relative dielectric constant  \cite{Chandra}. The electric dipolar energy drives ferroelectricity in some but not all cases. ``Improper" ferroelectricity  \cite{Improper} accompanies a structural phase transition in which electrostatics does not play a leading role. Since there is no unifying physical origin, many types of ferroelectrics exist. 

- \textit{Order-disorder ferroelectrics:} In materials with bistable protons or loosely locked molecular dipoles, the entropy increase with temperature drives a second-order phase transition that destroys a ferroelectric state at a critical temperature. This phenomenology is similar to magnetic systems and can be modeled by spin Hamiltonians  \cite{Blinc}.

- \textit{Displacive ferroelectrics:} Most ferroelectric phase transitions, including that of the iconic perovskite BaTiO\(_3\), are associated with an inversion symmetry-breaking structural phase transition triggered by a soft transverse optical phonon. The Landau-Devonshire theory based on a parameterized Landau free energy is an appropriate description for the ferroelectric phase transitions in these materials  \cite{Chandra}. 

- \textit{Shift-induced ferroelectricity} was discovered only recently in atomically thin bilayers of two-dimensional materials with hexagonal non-centrosymmetric unit cells such as  WTe\(_2\)  \cite{WTe2} and BN  \cite{Woods,Yasuda,Vizner}. Here a small slide or twist between the two monolayers reverses an electric dipole by mutual electric polarization and charge transfer. Landau theory \cite{Tang5} and first-principles calculations  \cite{Zhicheng} explain the observed high critical temperatures and low switching fields.

- \textit{Electronic ferroelectrics:}  Ionic displacements cause the permanent polarization in the above material classes. The inertia of the atomic masses slows down the switching dynamics. The reports of ferroelectricity caused by the ordering of electrons without rearrangement of the lattice  \cite{Fujiwara,Yamauchi,Moire} promise reduced response times. Even though the physics is very different, the phenomenology of polarization accumulation and transport is the same as for phonon-based ferroelectrics \cite{Adachi}

- \textit{Ferroelectric conductors:} According to conventional  wisdom electric conduction and ferroelectricity are incompatible because of the efficient screening of bulk charges by mobile electrons. Nevertheless, ferroelectric conductors and metal do exist in special materials with incomplete screening \cite{Polarmetals}. The perpendicular polarization in WTe\(_2\) \cite{WTe2}  or bilayer graphene  \cite{Moire}  can exist because two-dimensional conduction electrons cannot screen a permanent dipole oriented normal to the planes. A two-dimensional metal such as graphene also does not screen a homogeneous perpendicular polarization of a close-by ferroelectric \cite{Tang5}.  

- \textit{Antiferroelectrics:} The condensation of a Brillouin zone center (edge) phonon precedes the ferroelectric (antiferroelectric)  order. Ferrielectricity and more complex textures \cite{Randall} may arise during phase transitions that do not need to be displacive.   

- \textit{Multiferroics:} In multiferroic materials the order in the electric as well as magnetic polarizations coexist \cite{Meier}. ``Electromagnons'' are the excitations in multiferroic materials in the presence of ac electric fields that not necessarily carry electric dipoles  \cite{Electromagnons}. Electric and magnetic dynamics are coupled by the Moriya-Dzyaloshinskii interaction  \cite{Shen}.

\section{Ferronics}

Ferronics is the study of the elementary excitations of the ferroelectric order or
\textquotedblleft ferrons\textquotedblright  \cite{Bauer1,Tang1,Tang2,Tang3,Tang4,Tang5}, analogous with the subfield of magnonics in magnetism \cite{MagnonicsTB,Magnonics}. 

- \textit{Ferrons:} A ferron in state $i$ is a bosonic excitation that carries electric polarization in the form of a finite dipole $\mathbf{p}_{i}$. By a direct derivation  \cite{Tang2} or the Hellmann-Feynman theorem, the latter follows from the dependence of its frequency \(\omega_i=\epsilon_i/\hbar\) on an applied electric field $\mathbf{E}$
\begin{equation}
\mathbf{p}_{i}=-\frac{\partial\omega_{i}}{\partial\mathbf{E}},%
 \end{equation}
For comparison, the magnetic moment of a magnon \(j\) in simple ferromagnets with frequency \(\omega_j\)  reads  \(\mathbf{m}_{j}=-\partial\omega_{j}/ ( \mu_0 {\partial\mathbf{H}})\), where \(\mathbf{H}\) is the magnetic field and \(\mu_0\) the vacuum permeability. Its value is $m_{j}%
\approx-2\mu_{B},$ where $\mu_{B}$ is the Bohr magneton. A similar simple relation does not exist for the ferron polarization, however. 

The temperature dependence of the equilibrium polarization 
\begin{equation}
\Delta \mathbf{p}_{0}= \mathbf{p}_{0} (T)-\mathbf{p}_{0} (0)=  \sum_i \mathbf{p}_{i} f_P(\epsilon _i ,T),%
\end{equation}
in terms of the sum over the eigenstates of the system, where $f_P(\epsilon _i ,T)=\exp[\epsilon _i/(k_B T)-1]^{-1}$ is the Planck distribution and \(k_B\) Boltzmann's constant. The excitations of ferroelectrics (and ferromagnets) suppress the order, hence $p_{0}(T) \leq p_{0} (0)$ . 

The ferrons of most ferroelectrics are also phonons, but not all phonons are ferrons. The frequency changes of a molecular dipole of two oppositely charged ions in a parabolic potential and of optical phonons in polar dielectrics scale like $\delta \omega_{\mathrm{op}} \sim -E^{2}$. These systems, therefore, do not harbor ferrons in weak applied fields. Ferron excitations exist only as anharmonic oscillators in crystals with intrinsic or induced broken inversion symmetry, such as ferroelectrics or electrets. Ferron modes can be both transverse  (Goldstone-like) or longitudinal (Higgs-like), as well as acoustic or optical. 

- \textit{Models:}  In order-disorder ferroelectrics, the ferron dipole is caused by the reduced projection of the precessing dipole along the ferroelectric order or the hopping of protons. Both can be modeled by lattice dynamics  or spin models (P. Tang, in preparation) that access the full Brillouin zone \cite{Bauer1,Tang1}.  $\mathbf{p}_{i}$ depends strongly on band index and crystal momentum. It is mainly carried by optical modes, but in transport phenomena, the much larger group velocity of sound waves can compensate for their smaller dipoles. 

In displacive and shift-induced ferroelectrics the lattice vibrates in an inverted camelback potential that reduces the average dipole by longitudinal vibrations. Landau theory  \cite{Chandra} governs the ferron dispersion and mode amplitudes of displacive ferroelectrics in the long wavelength limit \cite{Tang2,Tang5}. 

In electronic ferroelectrics, ferrons emerge in the Falikov-Kimball model as excitations of an excitonic insulator phase  \cite{Adachi} . 

Equation (1) is a simple recipe to compute the ferron dipole by model or first-principles calculations. Many observables can be computed from the ferron energy dispersion \(\epsilon_{\nu \mathbf{k}}\) for a ferron band \(\nu\) and wave vector  \(\mathbf{k}\) and its polarization \(\mathbf{p}_{\nu \mathbf{k}}=-\partial\epsilon_{\nu \mathbf{k}}/ (\hbar\partial\mathbf{E})\) .

- \textit{Non-equilibrium:}  The equilibrium polarization \(\mathbf{p}_0\) density of a ferroelectric depends on temperature and the applied electric field. When the equilibrium of an isotropic medium is perturbed by gradients of temperature ($\partial_x T$)  and/or effective electric field (\(\partial_x E_\mathrm{eff}\)) parallel to the polarization in the \textit{x}-direction, the energy (\(j_q\)) and polarization (\(j_p\))  current densities flow to restore the equilibrium. When the perturbations are sufficiently small and slow in time and space, local linear response relations (Ohm's Law) apply. 
\begin{equation}
\left(
\begin{array}
[c]{c}%
-j_{p}\\
j_{q}%
\end{array}
\right)  =\sigma_p \left(
\begin{array}
[c]{cc}%
1 & S_p\\
\Pi_p & \kappa/\sigma_p
\end{array}
\right)  \left(
\begin{array}
[c]{c}%
\partial_x E_\mathrm{eff}\\
-\partial_x T
\end{array}
\right)  , \label{onsager}%
\end{equation}
where $\sigma_p$ $\left(  \kappa\right)  $ is the polarization (thermal) conductivity with units m/$\mathrm{\Omega}$ (WK\(^{-1}\)m\(^{-1}\)), while $S_p $ $\left(
\Pi_p =S_p T\right)  $ is the ferroelectric Seebeck (Peltier) coefficient with units VK\(^{-1}\)m\(^{-1}\) (V/m).  

- \textit{Coherent ferrons:} Similar to magnons, we can study ferrons in two different limits. Propagating magnons that are coherently excited by microwaves at GHz frequencies propagate over centimeters in high-quality magnets such as yttrium iron garnet (YIG). The ferrons in displacive ferroelectrics resonate at THz frequencies \cite{Zhuang}. Moreover, resonance line broadenings are governed by viscous dissipation that scales like $\sim \omega$, whence the lifetimes of optical phonons and their group velocities are relatively small. On the other hand, while their dipole and thus the coupling to ac and dc electric fields are relatively weak, acoustic ferrons are attractive for experiments on propagating ferrons in crystals with low acoustic attenuation and high group (sound) velocities. The dipolar interaction strongly affects ferrons at surfaces  \cite{TaoYu}. Excitation by a localized source leads to strongly focused ferrons beams on the surface of ferroelectrics. The strong interaction with the photons leads to substantial anticrossings between photon and ferron branches, i.e. ferron polaritons (P. Tang, in preparation).

- \textit{Transport:} In order to improve upon the simple diffusion theory sketched above, the transport coefficients in Eq. (1) can be computed from the ferron spectrum and dipoles. 

In constrictions such as point contacts, mean free paths may exceed the sample length. In that ballistic limit Landauer-B\"{u}ttiker scattering theory is appropriate  \cite{Tang1}. 

In the opposite diffuse regime, a semiclassical linearized Boltzmann equation leads to microscopic expressions for the parameters of the diffusion theory. When a non-equilibrium polarization or polarization accumulation (or ferron chemical potential) $\delta p = p(x,t)-p_0$ relaxes over a time $\tau_p$, the conservation relation for the polarization \(\partial_x j_{p}=- (\partial_t  - 1/\tau_p) \delta p\)
leads to a diffusion equation for \(\delta p\) that can be solved with appropriate boundary conditions\cite{Bauer1,Tang2}.   The effective field \(E_\mathrm{eff}\)  is the sum of applied and internal fields including a diffusion term proportional to \(\partial_x \delta p\). Its gradient is necessarily transient on a time scale \(\tau_p\), as is the field-driven thermal current or ``polarization Peltier effect". The polarization Seebeck effect is observable in terms of dc thermovoltages induced by a temperature gradient \cite{Bauer1,Tang1}. While both the Peltier effect and thermovoltage vanish with $\tau_p$, the polarization current induced by a temperature gradient persists and can be detected, e.g., by its stray magnetic fields  \cite{Tang1}. 

\section{Experiments}

Wooten \textit{et al}.   \cite{Wooten} reported the first experimental evidence of ferron excitations in terms of an electric-field dependence of the thermal (or heat) conductivity \(\kappa\) (Fig. 1) and velocity \(v\) of pressure waves in a lead zirconium titanate (PZT)-based ferroelectric device. The observed field dependence is consistent with the relation \(\kappa=(1/3) Cv^2\tau\), where the volumetric heat capacity at constant pressure  \(C\) is constant above the Debye temperature. Assuming that the scattering time  \(\tau\) does not depend on field \(d \ln \kappa /dE = 2 d \ln v/dE\) as observed.  These results indicate that the effect is intrinsic, \textit{i.e.} caused by the field modulation of acoustic phonons that carry an electric dipole. 

The original theories  \cite{Bauer1,Tang2} are not appropriate to model these results since (i) PZT is a displacive ferroelectric and (ii) in the Landau model the dipoles are carried exclusively by a single band of longitudinal soft optical phonons at high frequencies and small group velocities. The piezoelectric strain, \textit{i.e.} the linear contraction of the lattice by an applied electric field in ferroelectrics, modulates the sound velocities in particular via the mode and frequency-dependent Gr{\"u}neisen coefficients and thereby the thermal conductivity.  The model predicts that \(d \ln v/dE= -(d_{33}+2d_{31})\gamma\), where the piezoelectric coefficient \(d_{33}\) (\(d_{31}\)) measures the strain along (normal to) \textbf{E} and \( \gamma= d \ln \omega/d \ln V\) is the Gr¨uneisen constant, i.e. the derivative of the phonon frequency \(\omega\) with respect to the volume \(V\).   For the known material parameters for PZT the model explains the observations remarkably well. Below the Debye temperature  \(C \propto v^{-3}\)  and \(d \ln \kappa /dE = 2 d \ln v/dE\)  changes sign. Moreover, while \(d_{33}+2d_{31}>0\) in PZT, it is negative in Pb(Mg\(_{1/3}\)Nb\(_{2/3}\))O\(_3\)-PbTiO\(_3\) (PMN-PT), potentially enabling the design of complementary heat switches.

Preliminary data on non-ferroelectric materials with large piezoelectricity/electrostriction and high dielectric constants such as SrTiO\({_3}\) do not show any field dependence, which confirms the role of ferrons.  

Analogously, the modulation of the sound velocity in ferromagnetic materials  by magnetostriction  \cite{Deklerk}  may affect the thermal transport under an applied magnetic field. 

\begin{figure}[h]
	\centering
	\includegraphics{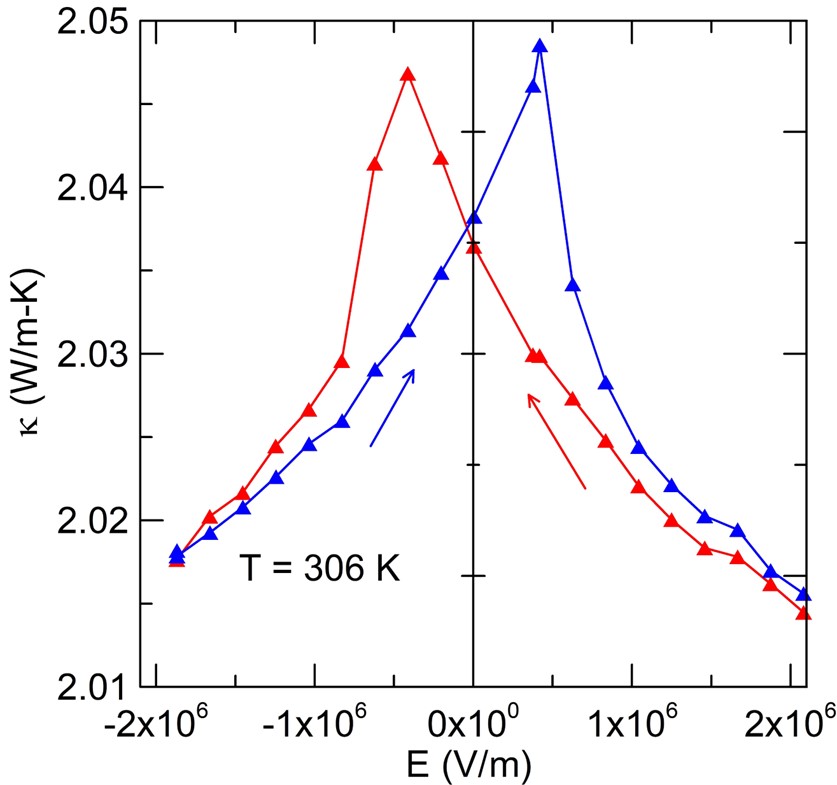}  
	\caption{Electric field dependence of the thermal conductivity of the ferroelectric PZT (adapted from \cite{Wooten} )}
\end{figure}

\section{Potential applications}

\textit{- Thermal management:} In the footsteps of spin caloritronics, the science and technology of coupled spin, heat, and charge currents \cite{ReviewSpinCal,SpinThermal}, ferrons offer new strategies for thermal control and power generation. 

Ferroelectric capacitors under temperature gradients generate stable surface charges, in contrast to the transient polarization offered by the pyroelectric effect. A transient polarization Peltier effect may contribute to electrocaloric cooling   \cite{Bauer1,Tang1}.

The polarization drag effect in heterostructures between ferroelectrics and metals frees device designers from the yoke of the Wiedemann-Franz Law. In such structures, heat and charge currents flow in parallel but are spatially separated in different materials. Choosing a ferroelectric with a low thermal and a metal with a high electric conductivity leads to respectable thermoelectric figures of merit in van der Waals stacks  \cite{Tang3}. In the spin Seebeck effect, the heat current in the magnet is also spatially separated from the induced transverse charge current in heavy-metal contacts, but the thermoelectric figure of merit is limited by relatively small spin Hall angles \cite{Cahaya},

The electric field-dependence of the thermal conductivity \cite{Wooten} could lead to electrically controlled all-solid-state heat switches between high- and low-conducting states with important potential applications \cite{Wehmeyer} in energy-saving technologies, particularly in engines with transient heating cycles such as solar-thermal power plants.  Their development would enable all-solid-state power generators and refrigeration technologies, improving the efficiency of electrocaloric and magnetocaloric engines.  Particularly interesting would be a set of complementary devices in a high- and low-conducting state that switches their role by an applied electric field that as well powers the active electrocaloric material, which is an option offered by the experiments \cite{Wooten}. 

\textit{- Logics and interconnects:} The weak dissipation of magnons makes them suitable conduits to exchange spin information and carry out logic operations \cite{SWcomputing}. Coherently excited ferrons can in principle carry out the same functions. The first theory of surface ferrons predicts new functionalities, such as tunable routing of focused ferron beams without the need for micro/nanostructuring the surface into wave guides  \cite{TaoYu}. Since spin currents are the bread and butter of spin-torque magnetic memories  \cite{STTMRAM}, polarization currents might be useful in the writing and read-out of ferroelectric memories  \cite{Memories}. 

\textit{- THz source:} When injected into a metal, the polarization current decays on a femtosecond scale. However, when injected into an undoped semiconductor with a fundamental band gap below thermal energies, the generated electron and holes may recombine to emit photons (G.E.W. Bauer, in preparation). When the non-radiative recombination times are sufficiently long, such a structure could operate as a THz radiation source driven by waste heat. 

\section{Conclusions and challenges}

The present Perspective only scratches the surface of a full understanding of the basic physics and the applications of electric-dipole carrying excitations in ferroelectrics and ferroelectric devices. Fundamental issues such as the formulation and observation of the polarization relaxation length must be still addressed. Some low-hanging fruits may be picked by applying the advanced insights of magnonics and mechanics to ferroelectric materials as in the  examples below.

\textit{- Charge-spin-polarization coupling:} While the spin-charge coupling is the bread and butter of spintronics, a polarization-charge coupling such as in the ferron drag effect \cite{Tang4}  is much less established. 

Ferroelectricity in a conducting channel strongly affects the spin-orbit interaction and switch spin textures in reciprocal space \cite{Rashba} and thereby the direction of spin Hall currents. Such a spin-polarization coupling should also work in the opposite direction, viz. creation of ferrons by spin currents. Natural or composite multiferroic materials have been the focus of many studies in the past decades, but the issues of polarization currents and their coupling to spin currents have to the best of our knowledge been overlooked. A systematic study of transport phenomena in the presence of polarization and magnetic order coupled by the Moriya-Dzyaloshinskii interaction  \cite{Shen} could lead, for instance, to polarization-charge coupling  current detectors. 

\textit{- Ferroelectric textures:} In unpoled ferroelectrics, the external electric fields are quenched by domain
with opposite polarizations that are separated by domain walls. Moving domain
walls facilitate ferroelectric switching by strongly suppressing critical fields. The interaction of ferrons with domain walls, which can be of the Ising type with vanishing polarization in
the domain wall center or characterized by a rotating order parameter similar
to ferromagnets \cite{Zhicheng} appears interesting and important. Ferroelectric domain walls may also provide waveguides for ferron excitations \cite{Xiao}. In analogy with magnonics, we envisage manipulation of ferroelectric textures by polarization currents, or in polar metals, even by charge currents.   

\textit{- Quantum effects:} Extending the techniques of optomechanical systems  \cite{Optomechanics} to ferroelectric materials would allow the quantum control of macroscopic electric polarization. 

In order to progress, more experiments are sorely needed. We hope that we have been able to inspire our colleagues and stimulate new collaborative efforts between experts in the fields of ferroelectrics and magnetism. 

\acknowledgements{}
P.T. and G.B. acknowledge the financial support by JSPS KAKENHI Grants No.~19H00645 and No.~22H04965. S.R.
was financially supported by Conselho Nacional de Desenvolvimento
Cient\'{\i}fico e Tecnol\'{o}gico (CNPq), Coordenac\~{a}o de Aperfeicoamento
de Pessoal de N\'{\i}vel Superior (CAPES), Financiadora de Estudos e Projetos
(FINEP), and Fundac\~{a}o de Amparo \'{a} Ci\^{e}ncia e Tecnologia do Estado
de Pernambuco (FACEPE). T.Y. acknowledges support by the National
Natural Science Foundation of China under Grant No.~0214012051, and the
startup grant of Huazhong University of Science and Technology (Grants
No.~3004012185 and No.~3004012198).  J.H. is supported by the U.S. National Science Foundation grant CBET-2133718 “Polarization Caloritronics: a pathway to electrically-controlled heat switches. 
R.I and K.U. are supported by JSPS KAKENHI Grants No. 20H02609 
and 22H04965 as well as JST CREST “Creation of Innovative Core Technologies for Nano-enabled Thermal Management” Grant No. JPMJCR17I1.

\end{document}